\documentclass[prl,twocolumn,preprintnumbers,amsmath,amssymb]{revtex4}

\usepackage{graphicx}
\usepackage{epsfig}
\usepackage{dcolumn}
\usepackage{bm}

\begin{document}

\title{Theory of the Stark Effect for P donors in Si}

\author{Mark Friesen}
\email{friesen@cae.wisc.edu}
\affiliation{Department of Physics, University of Wisconsin-Madison, Madison,
Wisconsin 53706-1390, USA }

\begin{abstract}
We develop an effective mass theory for substitutional donors in silicon in an
inhomogeneous environment. 
Valley-orbit coupling is included perturbatively. 
We specifically consider the Stark effect in Si:P.   
In this case, the theory becomes more accurate at high fields, 
while it is designed to give correct experimental binding energies 
at zero field.  Unexpectedly, 
the ground state energy for the donor electron is found to increase
with electric field as a consequence of spectrum narrowing of the 1$s$ manifold.  Our
results are of particular importance for the Kane quantum computer.
\end{abstract}


\maketitle

Phosphorus doped silicon is one of the most well-studied semiconducting systems
\cite{yubook}, owing to its importance in the electronics 
industry.  A more recent and exotic application is quantum computing, in which
the Kane proposal posits nuclear spin qubits on Si:P, with interactions
mediated by donor-bound electrons \cite{kane98}.  By tuning the potentials on 
aligned electrodes, the electrons are brought to the point of ionization, 
in order to control
nuclear hyperfine interactions and the overlap between neighboring qubits.
Spin dependent ionization also provides a means for electrical detection of the
qubit state.  The precise control of donor-bound electrons in such a
complex environment remains an experimental challenge \cite{dzurak}, 
requiring detailed theoretical input \cite{fang02,kettle03,smit03}.  

The theory of isolated donors in silicon remains one of the crowning achievements of
solid-state physics.  One of the most effective treatments for shallow donors 
is the effective mass approximation (EMA) \cite{kohn}, in which 
the donor potential is assumed to vary slowly compared to the crystal 
potential.  As a result, the long and short-wavelength physics decouple.  An excess
electron on the donor can be described by an envelope equation --
a Schr\"{o}dinger equation with an effective mass and a dielectric constant.  The 
theory highlights the most essential feature of 
silicon's bulk band structure:  the six-fold degeneracy of the conduction valleys.
Near the impurity core, the assumption of slowly varying potential breaks down,
causing valley-orbit interactions to couple envelope functions in different 
valleys.  A careful treatment of the potential very near the impurity (the 
`central cell' region) enables estimates for the energy splitting of the six valley 
states, which are in good agreement with experiments \cite{ramdas81}.

In a more general, inhomogeneous environment, silicon donors can be studied by
tight-binding techniques \cite{martins04}.  However, existing EMA theories 
introduce severe approximations that limit their scope. 
Many important questions therefore remain open.  For example, there
is no theory to determine how the weight of a donor wavefunction will  
redistribute among the six conduction valleys in the presence of a generic (i.e.,
low symmetry) potential.  Since the envelope functions in different 
valleys have different energies (due to anisotropy of the effective mass), 
this is an important question.

In the context of quantum computing, several recent papers obtain tractible results
for donor 
ionization, by ignoring valley-orbit interactions \cite{fang02,kettle03,smit03}.
This single-valley approach provides a useful picture of  
the distorted envelopes in the individual valleys.  
However, it does not capture the spectrum narrowing of the 1$s$ manifold.
Smit \textit{et al.} have recently studied the Stark effect, beyond 
the single-valley EMA, 
by applying symmetry arguments and perturbation theory \cite{smit04}.
However, their results are only applicable at low fields, and they also
do not capture spectrum narrowing.  In the present Letter, we show how to overcome
such difficulties.  We develop a multi-valley EMA for shallow donors in silicon
in a general, inhomogeneous environment.  

\begin{figure}[t]
\centerline{\epsfxsize=2.5in \epsfbox{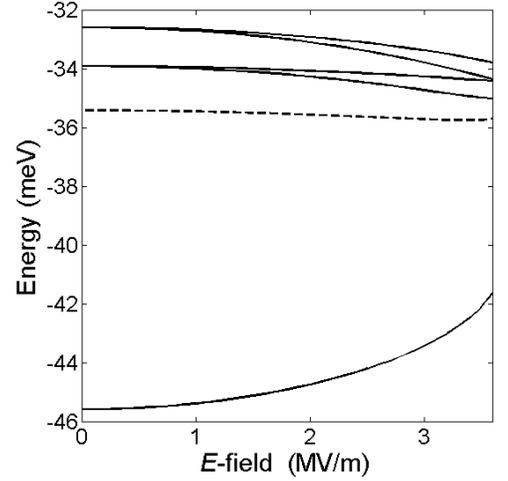}}
\caption{Solid curves:  the variational binding energy as a function of electric 
field for the 1$s$ manifold in Si:P.  
(The second solid curve from bottom is doubly degenerate.)
Dashed curve:  average energy of the manifold.  
\label{fig:energy}}
\end{figure}

Before outlining the theory, we discuss two competing effects that 
determine the energy shift of electrons in the 1$s$ manifold.  
The most well known effect is the quadratic Stark shift, which causes the 
\textit{average} energy
of the six valleys to decrease with field.  The second effect
is spectrum narrowing within the 1$s$ manifold.  Valley-orbit effects induce 
a manifold splitting of about 13~meV for the six 1$s$ states in Si:P.  
As the field increases, the donor electron is gradually pulled 
off the impurity and out of the central cell, causing the splitting to narrow.  
Consequently, 
the ground state shifts upwards toward the average energy of the manifold.  An 
accurate, quantitative analysis is needed to establish the dominant effect.  
The energies computed in our EMA analysis are shown in Fig.~\ref{fig:energy}.
We find that spectrum narrowing is the dominant effect for the ground state, causing
the energy to increase with field.  
A similar result was obtained recently in tight binding simulations of
a gated qubit \cite{martins04}.  However, the differences in that geometry make  
comparison difficult.  Here, the surprising behavior arises directly from the 
multi-valley band structure.

We now describe the effective mass theory.  The wavefunction of the excess electron of
the donor impurity can be written as \cite{kohn}
\begin{equation}
\Psi({\bm r}) = \sum_{i=1}^{6} \alpha_i \phi_i({\bm r}) F_i({\bm r}) , \label{eq:EMTpsi}
\end{equation}
where $\alpha_i$ are the valley composition parameters, reflecting the portion of the
wavefunction in each of the six valleys.  Normalization gives 
$\sum_{i=1}^{6} |\alpha_i|^2 = 1$.  The Bloch functions are given by 
$\phi_i ({\bm r}) = u_i ({\bm r}) e^{i {\bm k}_i\cdot {\bf r}}$, 
where ${\bf k}_i=k_0\hat{i}$ specifies the minimum of the $i$th conduction valley,
and $u_i$ has the periodicity of the crystal lattice.  The $F_i$ are envelope 
functions for the six valleys.  

To develop an envelope equation for $F_i$, we consider the potential energy 
$U({\bm r})$, including both the impurity $V_i({\bm r})$ and its surroundings, but
excluding the crystal potential \cite{kohn}.  In this Letter, we specifically consider
a uniform external field $\bm{E}=E\hat{z}$.  However, the results are easily 
generalized.  The impurity ion is not a perfect point charge.  Deviations from point 
charge behavior $U_\text{cc}({\bm r})$ are called `central cell corrections,' because
they are strongly localized within a central cell radius of 1-2~\AA\ \cite{ning71}.  
Thus, $U({\bm r})= V_i({\bm r})-eEz = [-e^2/4\pi\epsilon r+U_\text{cc}({\bm r})]-eEz$.
Central cell effects can also include the breakdown of the concept of the 
dielectric constant and local distortions of the Si lattice near the P ion.  
Although there has been progress in the analytical description of the central cell 
\cite{pantelides}, a full elucidation is challenging.  Below, we overcome this
difficulty by introducing a semi-analytical treatment of central cell effects, using
the known energy spectrum of the Si:P donor electron at zero field.

The envelope equation of Fritzsche and Twose can be extended to include an external
field.  The equation goes beyond the single-valley
EMA by including valley-orbit interactions \cite{pantelides,wilson61,fritzsche,twose}:
\begin{eqnarray}
\sum_{j=1}^{6} \alpha_j e^{i {\bm k}_j \cdot {\bm r}}
[T_j(-i\hbar {\bm \nabla}) -\frac{e^2}{4\pi\epsilon r}-eEz-\varepsilon ] F_j({\bm r})
&& \nonumber \\
+ \sum_{j=1}^{6} \alpha_j e^{i {\bm k}_j \cdot {\bm r}}
U_\text{cc}({\bm r}) F_j({\bm r})=0  .\quad \quad  && \label{eq:sah}
\end{eqnarray}
We refer to the first line of (\ref{eq:sah}) as EMA terms, and the second line as
central cell corrections.  Because the central cell terms are 
localized so strongly near the impurity, $U_\text{cc}({\bm r})$ is essentially a 
contact potential \cite{twose}.  This allows us to replace $F_j({\bm r})$ 
by $F_{0j}$, the envelope amplitude at the impurity site.

In the so-called multi-valley EMA of Eq.~(\ref{eq:sah}), 
$\varepsilon$ is the energy eigenvalue. 
The kinetic energy operator $T_i({\bm k})$ is the quadratic expansion of the 
conduction band dispersion relation $E_c( {\bm k})$ around the valley minimum 
${\bm k}_i$.  This prescription for $T_i$ introduces spurious inter-valley  
contributions, which can be avoided if necessary \cite{shindo76,resta77}.
However, in our theory, only the intra-valley energy expression 
will be computed explicitly.  For this case, the quadratic expansion is appropriate. 
Note that we have limited our 
consideration to the six low-lying valleys in the 1$s$ manifold, 
as appropriate for the EMA \cite{kohn}.  
Higher bands give contributions much smaller than the terms considered here.  

To solve the envelope equation we take a perturbative approach by assuming that valley-orbit
interactions are weak.  We shall justify this assumption later, by comparing the sizes of
various terms.  At zeroth order in the perturbation, we solve the uncoupled, 
single-valley equations to obtain the eigenfunctions 
$F_j^{(0)}({\bm r})$.  These are then used to compute the valley-orbit terms.
The resulting first-order Hamiltonian is given
in the valley basis $(+x,-x,+y,-y,+z,-z)$:
\begin{equation}
H=\begin{pmatrix}
\tilde{E}_x&\Delta_{1x}&\Delta_{2xy}&\Delta_{2xy}&\Delta_{2xz}&\Delta_{2xz}\\
\Delta_{1x}&\tilde{E}_x&\Delta_{2xy}&\Delta_{2xy}&\Delta_{2xz}&\Delta_{2xz}\\
\Delta_{2xy}&\Delta_{2xy}&\tilde{E}_x&\Delta_{1x}&\Delta_{2xz}&\Delta_{2xz}\\
\Delta_{2xy}&\Delta_{2xy}&\Delta_{1x}&\tilde{E}_x&\Delta_{2xz}&\Delta_{2xz}\\
\Delta_{2xz}&\Delta_{2xz}&\Delta_{2xz}&\Delta_{2xz}&\tilde{E}_z&\Delta_{1z}\\
\Delta_{2xz}&\Delta_{2xz}&\Delta_{2xz}&\Delta_{2xz}&\Delta_{1z}&\tilde{E}_z
\end{pmatrix} , \label{eq:Hvo}
\end{equation}
where $\tilde{E}_i=E_i+\Delta_{0i}$, and we have used axial 
($\tilde{E}_x=\tilde{E}_y$) and time reversal ($\tilde{E}_{+z}=\tilde{E}_{-z}$) 
symmetry.  The various terms in $H$ are defined as follows:
\begin{equation}
E_j = \int {F^{(0)}_j}
[T_j-\frac{e^2}{4\pi\epsilon r}-eEz]F^{(0)}_j \, d^3r , \label{eq:Ej} 
\end{equation} \vspace{-.2in} \begin{equation}
\Delta_{0j} = \int {F^{(0)}_j} U_\text{cc} F^{(0)}_j \, d^3r \label{eq:Delta0} 
\end{equation} \vspace{-.2in} \begin{equation}
\Delta_{1j} = \int {F^{(0)}_j} e^{-2i{\bm k}_j\cdot {\bm r}}
U_\text{cc}F^{(0)}_j \, d^3r  \label{eq:Delta1}
\end{equation} \vspace{-.2in} \begin{equation}
\Delta_{2ji} = \int {F^{(0)}_j} e^{i({\bm k}_i-{\bm k}_j)\cdot {\bm r}}
U_\text{cc}{F^{(0)}_i} \, d^3r \quad (i\neq j) . \label{eq:Delta2} 
\end{equation}

Equation~(\ref{eq:sah}) also generates off-diagonal terms in the Hamiltonian, 
analogous to (\ref{eq:Ej}).
Because of the fast oscillations in the off-diagonal integrals and the smoothness of 
the effective mass eigenfunctions at zeroth order, such terms can be ignored 
in comparison with $E_j$ \cite{fritzsche}.  The only non-vanishing 
off-diagonal terms are $\Delta_{1j}$ and $\Delta_{2ji}$, due to the strong localization
of $U_\text{cc}$ on atomic length scales.  
Approximating $U_\text{cc}$ as a contact potential \cite{twose} gives
\begin{equation}
\Delta_{0j} = v_0F_{0j}^2, \quad 
\Delta_{1j}=v_1F_{0j}^2, \quad 
\Delta_{2ji}=v_2F_{0j}F_{0i}, \label{eq:firstorder}
\end{equation}
where $F_{0j}$ is the magnitude of the envelope function at the impurity site.  Note
that we have chosen to include the electric field in the $E_j$ terms, rather than
the central cell terms.  This assignment is justified by comparing $-eEz$ with the
point charge potential $-e^2/4\pi\epsilon r$.  Within the central cell, the former
becomes dominant only for fields much larger than the ionization field.  The
contact potentials $v_{0,1,2}$ are therefore independent of field.  They can be determined 
empirically at
zero field, where the energy spectrum of the donor is well known.  This is a central
observation of our theory.  Once $v_{0,1,2}$ are specified, the field dependence of the
valley-orbit terms arises only from the wavefunction normalization
$F^{(0)}_{0j}$.  Since $F^{(0)}_{0j}$ decreases with increasing field, valley splitting also
decreases.  This is the origin of spectrum narrowing.  

We now perform the Stark shift calculation for Si:P using the pertubation theory.  
The unperturbed (single-valley) envelope equations are given by
\[
\left[ -\frac{\hbar^2}{2m_l} \frac{\partial^2}{\partial x^2}
-\frac{\hbar^2}{2m_t} \left( \frac{\partial^2}{\partial y^2}
+\frac{\partial^2}{\partial z^2} \right)
-\frac{e^2}{4\pi\epsilon r} -eEz \right] F^{(0)}_x (\bm{r}) 
\] \vspace{-.25in}
\begin{equation}
= E^{(0)}_x F^{(0)}_x ({\bm r}) ,\label{eq:envFx} 
\end{equation} \vspace{-.3in}
\[
\left[ -\frac{\hbar^2}{2m_l} \frac{\partial^2}{\partial z^2}
-\frac{\hbar^2}{2m_t} \left( \frac{\partial^2}{\partial x^2}
+\frac{\partial^2}{\partial y^2} \right)
-\frac{e^2}{4\pi\epsilon r} -eEz \right] F^{(0)}_z (\bm{r}) 
\] \vspace{-.25in}
\begin{equation}
= E^{(0)}_z F^{(0)}_z ({\bm r}) .\label{eq:envFz}
\end{equation}
The equation for $F^{(0)}_y({\bm r})$ is identical to $F^{(0)}_x({\bm r})$, 
with $x\leftrightarrow y$.
The longitudinal and transverse effective masses for silicon are 
$m_l^*=0.916m_0$ and $m_t^*=0.191m_0$, respectively.  The low temperature dielectric
constant is $\epsilon =11.4\epsilon_0$.

We use a variational method to solve Eqs.~(\ref{eq:envFx}) and (\ref{eq:envFz}),
obtaining good agreement with numerical results.  The variational forms are given by 
\[
F^{(0)}_x({\bm r}) = F_{0x} (1+q_xz) \exp 
\left(-\sqrt{x^2/a_x^2+(y^2+z^2)/b_x^2} \right), \] \vspace{-.2in} \[
F^{(0)}_z({\bm r}) = F_{0z} (1+q_zz) \exp 
\left(-\sqrt{z^2/a_z^2+(x^2+y^2)/b_z^2} \right), 
\]
These functions are similar to ones used
successfully in the zero-field case \cite{kohn}.
At this order, the envelope equations are uncoupled, and the valley energies
$E^{(0)}_{x,z}$ are minimized independently.  

The contact potentials are determined at zero field.  
In this limit, the problem is isotropic,
giving $E^{(0)}_x=E^{(0)}_z=-31.28$~meV, 
$a_x=a_z=1.360$~nm, $b_x=b_z=2.365$~nm, $q_x=q_z=0$,
and $F_{0x}=F_{0z}=6.469\times 10^{12}$~m$^{-3/2}$.  
The central cell terms do not require valley indices:  $\Delta_0, \Delta_1, \Delta_2$.
The zero-field binding energies are well known from experiments
\cite{ramdas81}:  $\epsilon =-45.59 (1), -33.89 (3),$ and $-32.58 (2)$~meV (parentheses
indicate the zero-field degeneracies).  By requiring the Hamiltonian (\ref{eq:Hvo})
to provide these eigenvalues, the matrix elements are uniquely specified:
$\tilde{E}=-35.40$~meV, $\Delta_0=-4.13$~meV, $\Delta_1=-1.51$~meV, $\Delta_2=-2.17$~meV, 
leading to the identification
$v_0=-1.58\times 10^{-47}$~Jm$^3$, $v_1=-5.78\times 10^{-48}$~Jm$^3$, 
and $v_2=-8.31\times 10^{-48}$~Jm$^3$, from Eq.~(\ref{eq:firstorder}).  
Note that these values depend on our choice of variational functions.
We can now justify the perturbation approach by comparing the perturbation terms
$\Delta_{0,1,2}$ to the
EMA energy $E^{(0)}$.  The theory improves in accuracy with increasing 
field, as the central cell terms grow smaller. 
However, the method is also designed to obtain exact experimental results for 
the donor binding energies at zero-field.  

At non-zero fields, we use the same
perturbation prescription.  The variational form 
is used to minimize the energies of the uncoupled envelope functions, 
giving $E^{(0)}_{x,z}$ and
$F^{(0)}_{x,z}$.  The off-diagonal elements of $H$ are obtained from 
(\ref{eq:firstorder}), 
using the field-independent values of $v_{0,1,2}$ just obtained.  

We diagonalize $H$ to obtain the first-order eigenstates.  
The resulting eigenvectors are the valley composition parameters $\alpha_i$.
Using the zero-field notation of Ref.~\cite{smit04}, we obtain
the lowest ($g$) and highest ($r$) eigenvalues:
\begin{eqnarray}
\varepsilon_{g,r} \!\! &=& \! \frac{1}{2}\Big[ \tilde{E}_x+\tilde{E}_z+\Delta_{1x}+
\Delta_{1z}+2\Delta_{2xy}
\\ && \!
\pm \sqrt{32\Delta_{2xz}^2+(\tilde{E}_x-\tilde{E}_z+\Delta_{1x}-\Delta_{1z}+
2\Delta_{2xy})^2} \Big] .
\nonumber
\end{eqnarray}
Corresponding eigenvectors are expressed in the valley basis:
\begin{equation}
{\bm \alpha}_{g,r} = \Big(1,1,1,1,{\alpha'}_{g,r},{\alpha'}_{g,r}\Big)/
\sqrt{4+2{{\alpha'}_{g,r}}^2} , \label{eq:epg}
\end{equation}
where
\begin{eqnarray}
{\alpha'}_{g,r} \!\! &=& \! \frac{1}{4\Delta_{2xz}}
\Big[ -(\tilde{E}_x-\tilde{E}_z+\Delta_{1x}-\Delta_{1z}+2\Delta_{2xy}) 
\nonumber \\ && \!\!\!
\pm \sqrt{32\Delta_{2xz}^2+(\tilde{E}_x-\tilde{E}_z+\Delta_{1x}-\Delta_{1z}
+2\Delta_{2xy})^2} \Big]
. \nonumber \\
\end{eqnarray}
At zero field, we recover the expected \cite{kohn} eigenstates
${\bm \alpha}_g=(1,1,1,1,1,1)/\sqrt{6}$ and ${\bm \alpha}_r=(1,1,1,1,-2,-2)/\sqrt{12}$.

The remaining eigenvectors are the same as their zero-field expressions:
\begin{eqnarray}
{\bm \alpha}_x &=& (1,-1,0,0,0,0)/\sqrt{2} ,\\
{\bm \alpha}_y &=& (0,0,1,-1,0,0)/\sqrt{2} ,\\
{\bm \alpha}_z &=& (0,0,0,0,1,-1)/\sqrt{2} ,\\
{\bm \alpha}_s &=& (1,1,-1,-1,0,0)/2 ,
\end{eqnarray}
with eigenvalues, from lowest to highest:
\begin{equation}
\epsilon_{x,y} = \tilde{E}_x-\Delta_{1x} ,
\end{equation} \vspace{-.25in} \begin{equation}
\epsilon_z = \tilde{E}_z-\Delta_{1z} ,
\end{equation} \vspace{-.25in} \begin{equation}
\epsilon_s = \tilde{E}_x+\Delta_{1x}-2\Delta_{2xy} .
\end{equation}

\begin{figure}[t]
\centerline{\epsfxsize=2.5in \epsfbox{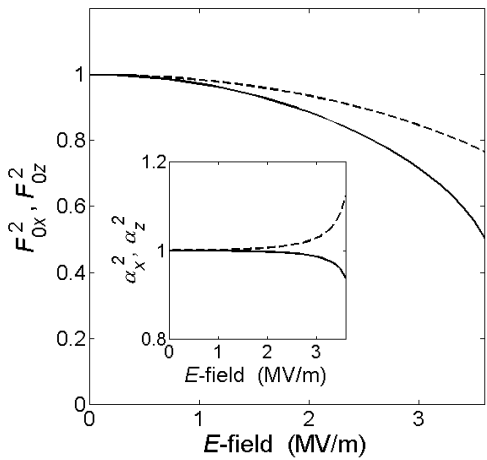}}
\caption{
Amplitude of the ground state envelope function 
vs.\ electric field.  Inset:  ground state valley composition parameters.
The solid lines correspond to the $x$ valley.
The dashed lines correspond to the $z$ valley.
All curves are normalized to 1 at zero field.
\label{fig:F0alpha}}
\end{figure}

The resulting energies, obtained from this perturbation-variational theory, are plotted 
in Fig.~\ref{fig:energy}.  For increasing fields,
the wavefunction moves off the impurity site, as shown in Fig.~\ref{fig:F0alpha}.  
In both figures, the results are displayed for fields up to the critical field 
($\sim 3.7$~MV/m), beyond which the system is completely unstable to ionization,
and solutions cannot be obtained.
Because of the observed upturn of the ground state energy with electric field, 
ionization occurs at considerably lower fields than expected from a single-valley EMA.
The latter predicts a downturn of $\epsilon_g$ with field.  Clearly multi-valley
effects play an crucial role in ionization calculations.

An interesting question is whether the electric
field can redistribute the weight of the electron between the six valleys.  As shown in
the inset of Fig.~\ref{fig:F0alpha}, the answer is ``yes."  The effect is small,
except near the critical field.  However, for many donor-bound qubit schemes, 
the ionized or nearly-ionized donor states are utilized for gate operations
\cite{kane98,vrijen00,skinner03}, so the regime is of practical interest.
Valley redistribution is also of interest when an unintended 
impurity interacts with a quantum dot qubit in a quantum well \cite{note}.  
Because of strain effects, the qubit wavefunction comprises only the two $z$ 
valleys \cite{boykin04}, so an exchange coupling between the qubit and impurity 
electrons occurs 
primarily in the $z$ valley channel.  It is therefore necessary to know what
portion of the donor wavefunction resides in the $z$ valleys.

In conclusion, we have developed a multi-valley EMA for donor-bound electrons in silicon 
in an inhomogeneous potential.  The theory is applied to the Stark shift in Si:P.  
In contrast with previous theories, we predict that the 
ground state energy of the donor will increase with electric field, due to spectrum narrowing of the 1$s$ manifold.  Comparison with previous results \cite{smit04}
gives corrections as large as 10~meV.  Such theories also do not address the
valley redistribution of the wavefunction, which can amount to $15\%$.  
The new effects are most
prevalent in the ionization regime, where quantum computers are expected to operate.
These remarkable results show that even qualitative descriptions of 
shallow donors in Si must take into account valley-orbit interactions.  

\begin{acknowledgments}
I acknowledge many useful conversations with C. Tahan, R. Joynt, M. Eriksson,
and S. Liao.  This work was supported by NSA and ARDA under ARO contract  
number W911NF-04-1-0389, and by the National Science Foundation through the ITR 
program (DMR-0325634) and the QuBIC program (EIA-0130400).
\end{acknowledgments}


\begin{thebibliography}{99}

\bibitem{yubook}
P. Y. Yu and M. Cardona, \textit{Fundamentals of Semiconductors: 
Physics and Materials Properties}, 3rd ed. (Springer-Verlag, Berlin, 2001).

\bibitem{kane98}
B. E. Kane, Nature (London) {\bf 393}, 133 (1998).

\bibitem{dzurak}
A. S. Dzurak, \textit{et al.}, cond-mat/0306265 (unpublished).

\bibitem{fang02}
A. Fang, Y. C. Chang, and J. R. Tucker, \prb \textbf{66}, 155331 (2002).

\bibitem{kettle03}
L. M. Kettle, \textit{et al.} \prb \textbf{68}, 075317 (2003).

\bibitem{smit03}
G. D. J. Smit, S. Rogge, J. Caro, and T. M. Klapwijk, \prb \textbf{68}, 193302 (2003).

\bibitem{kohn}
W.~Kohn, in \textit{Solid State Physics}, edited by F.~Seitz and 
D.~Turnbull (Academic Press, New York, 1957), Vol.~5.

\bibitem{ramdas81}
For a review, see
A. K. Ramdas and S. Rodriguez, Rep. Prog. Phys. \textbf{44}, 1297 (1981).

\bibitem{martins04}
A. S. Martins, R. B. Capaz, and B. Koiller, \prb \textbf{69}, 085320 (2004).

\bibitem{smit04}
G. D. J. Smit, S. Rogge, J. Caro, and T. M. Klapwijk, \prb \textbf{70}, 035206 (2004).

\bibitem{ning71}
T. H. Ning and C. T. Sah, \prb \textbf{4}, 3468 (1971).

\bibitem{pantelides}
S. T. Pantelides and C. T. Sah, \prb \textbf{10}, 621 (1974).

\bibitem{wilson61}
D. K. Wilson and G. Feher, Phys. Rev. \textbf{124}, 1068 (1961).

\bibitem{fritzsche}
H. Fritzsche, Phys. Rev. \textbf{125}, 1560 (1962).

\bibitem{twose}
W. D. Twose, in the Appendix of Ref.~\cite{fritzsche}.

\bibitem{shindo76}
K. Shindo and H. Nara, J. Phys. Soc. Japan \textbf{40}, 1640 (1976).

\bibitem{resta77}
R. Resta, J. Phys. C \textbf{10}, L179 (1977).

\bibitem{vrijen00}
R. Vrijen, E. Yablonovitch, K. Wang, H. W. Jiang, A. Balandin, V. Roychowdhury, T. Mor, and D. DiVincenzo, \pra \textbf{62}, 012306 (2000).

\bibitem{skinner03}
A. J. Skinner, M. E. Davenport, and B. E. Kane, \prl \textbf{90}, 087901 (2003).

\bibitem{note}
This problem motivated the current study.  See
M. Friesen, \textit{et al.}, \prb \textbf{67}, 121301(R) (2003);
S. Liao and M. Friesen (unpublished).

\bibitem{boykin04}
T. Boykin, \textit{et al.}, Appl. Phys. Lett. \textbf{84}, 115 (2004).

\end{thebibliography}
\end{document}